# Axion Bose-Einstein Condensation: a model beyond Cold Dark Matter

Q. Yang

*Department of Physics, University of Florida, Gainesville, Florida 32611, USA*

**Abstract.** Cold dark matter axions form a Bose-Einstein condensate if the axions thermalize. Recently, it was found [1] that they do thermalize when the photon temperature reaches T ~ 100 eV$(f/10^{12}\text{GeV})^{1/2}$ and that they continue to do so thereafter. We discuss the differences between axion BEC and CDM in the linear regime and the non-linear regime of evolution of density perturbations. We find that axion BEC provides a mechanism for the production of net overall rotation in dark matter halos, and for the alignment of cosmic microwave anisotropy multi-poles.

"Cold dark matter" (CDM) is widely believed an important part of the universe based on present observations and calculations. Although ordinary baryonic dark matter is present as well, most of dark matter is non-baryonic weakly interacting particles. The word "cold" indicates the particles have very small primordial velocity dispersion.

The axion was first suggested to solve the "strong CP problem" of QCD [2]. Later, experiment and theoretical considerations gave it a mass window: $3*10^{-3}$eV-$10^{-6}$eV [3] [4]. There are two cosmological axion populations: thermal axions are created through interactions among particles in the primordial soup and cold axions are created when the axion mass turned on at QCD time:

$$t_1 \approx 2 \cdot 10^{-7} \sec(f/10^{12}\text{GeV})^{1/3}.$$

The cold axions have small velocity dispersion. If they are completely interaction-less, then their velocity dispersion is:

$$\Delta v \sim \frac{a(t_1)}{a(t)} \cdot \frac{1}{mt_1} \qquad (1)$$

where *a(t)* is the cosmological scale factor and *m* is the axion mass. The velocity dispersion determines the effective temperature of cold axions. In case inflation occurs after the Peccei-Quinn phase transition, $\Delta v$ is even smaller because the axion field gets homogenized during inflation [5].

The number density of cold axions is:

$$n(t) \sim \frac{4 \cdot 10^{47}}{\text{cm}^3}(\frac{f}{10^{12}\text{GeV}})^{5/3}(\frac{a(t_1)}{a(t)})^3. \qquad (2)$$

Here, $f$ is the axion decay constant. It is related to the axion mass by:

$$m \approx 6 \cdot 10^{-6} \text{eV} \frac{10^{12} \text{GeV}}{f}. \tag{3}$$

Clearly, the average phase space density is very high due to the combination of high number density and small velocity dispersion. Boson-Einstein condensation requires the particles to have a conserved charge. Indeed, the formation of BEC is due to the impossibility of adding additional charges in the thermally excited states, so the ground state stores the exceeding charges. The conserved charge is axion number in this case. The number of axions is not exactly conserved, since, for example, the axion can decay to two photons due to the interaction:

$$L_{a\gamma\gamma} = -g_{\varphi\gamma\gamma} \varphi(x) \vec{E} \cdot \vec{B} \tag{4}$$

where $\varphi$ is the axion field and $g_{\varphi\gamma\gamma}$ is the photon-axion coupling constant. However, due to the extreme weak coupling, the decay rates of axion number changing processes are very small compared with the Hubble rate, which means effectively axion particle number is a conserved charge. The critical temperature to form axion BEC is:

$$T_c = \left(\frac{\pi^2 n}{\zeta(3)}\right)^{1/3} \approx 300 \text{GeV} (f/10^{12} \text{GeV})^{5/9} \frac{a(t_1)}{a(t)} \tag{5}$$

which is much higher than the effective temperature of cold axions.

According to the discussion above, the only condition for axion BEC that is not manifestly satisfied is thermal equilibrium. Axions are in thermal equilibrium if their relaxation rate $\Gamma$ is much larger than the Hubble expansion rate. Thermal equilibrium seems unlikely because the axion is very weakly coupled. The self-interaction term is $(\lambda/4!)\varphi^4$, with $\lambda \approx 0.35(m^2/f^2)$. The $\varphi + \varphi \rightarrow \varphi + \varphi$ scattering cross-section is:

$$\sigma_0 = 1.5 \cdot 10^{-105} \text{cm}^2 (m/10^{-5} \text{eV}). \tag{6}$$

The highly occupied final quantum states give an enhancement factor $N$ to the scattering process [6], so the relaxation rate is:

$$\Gamma \sim n\sigma_0 \delta v N \tag{7}$$

where $n$ is physical space density, $\delta v$ is velocity dispersion and $N$ is phase space density. Eq.(7) is valid only when the energy dispersion is large compared to the relaxation rate, a necessary condition for the validity of the random phase approximation. Shortly after $t_1$, the condition is no longer satisfied and the relaxation rate is $\Gamma \sim \lambda n / m^2$ instead. One find that the self-interactions barely thermalize axions at QCD time $t_1$ and stop to do so afterward.

It turns out the gravitational interactions, becoming more and more important as the horizon expands, play the major role in thermalize axions. We find the gravitational interactions thermalize the axions at the rate:

$$\Gamma_g \sim Gnm^2 l^2 \tag{8}$$

where the correlation length $l$ is:

$$l \sim \frac{1}{m\Delta v}. \tag{9}$$

The gravitational interactions thermalize the axions after photon temperature T < 100eV$(f/10^{12}$ GeV$)^{1/2}$ and continue to do so afterward .

Because the axions continue to thermalize, the majority of particles will be in the lowest energy available state. To compare axion BEC and CDM we divide the observations into three arenas: 1) the linear regime (1st order of perturbation theory) of evolution within the horizon; 2) the non-linear regime; 3) the behavior on the scale of the horizon.

The axion field satisfies the Heisenberg equation of motion:

$$D^\mu D_\mu \varphi(x) = g^{\mu\nu}[\partial_\mu \partial_\nu - \Gamma^\lambda{}_{\mu\nu}\partial_\lambda]\varphi(x) = m^2 \varphi(x). \tag{10}$$

We have neglected the self-interaction term $(-1/6)\lambda\varphi^3$ which is only important for a very short period after QCD time. The axion field can be expanded in particle modes as:

$$\varphi(x) = \sum [a_{\vec{\alpha}} \phi_{\vec{\alpha}}(x) + a_{\vec{\alpha}}^\dagger \phi_{\vec{\alpha}}^*]. \tag{11}$$

Except for a tiny fraction, most axions go to a single state that we label as $\vec{\alpha} = 0$. $\phi_0(x)$ is the corresponding wave function. The state of the axion field is: $|N\rangle = (1/\sqrt{N!})(a_0^+)^N |0\rangle$ where $|0\rangle$ is the vacuum and N is the particle number. In the spatially flat homogeneous and isotropic Robertson-Walker space-time,

$$\phi_0 = \frac{A}{a(t)^{3/2}} e^{-imt} \tag{12}$$

where A is a constant. The stress-energy-momentum tensor has expectation value:

$$\langle N|T_{\mu\nu}|N\rangle = N[\partial_\mu \phi_0 \partial_\nu \phi_0^* + \partial_\nu \phi_0 \partial_\mu \phi_0^* + g_{\mu\nu}(-\partial_\lambda \phi_0 \partial^\lambda \phi_0^* - m^2 \phi_0 \phi_0^*)]. \tag{13}$$

In a flat Minkowski space-time and in the non-relativistic limit, neglecting terms of order $(1/m)\cdot\partial_t$ compared to terms of order one, Eq.(10) implies the Schrödinger equation.

$$i\partial_t \Psi = -\frac{\nabla^2}{2m}\Psi. \tag{14}$$

The wave function can be written as [7]:

$$\Psi(\vec{x},t) = \frac{1}{\sqrt{2mN}} B(\vec{x},t) e^{i\beta(\vec{x},t)}. \tag{15}$$

In terms of $B(\vec{x},t)$ and $\beta(\vec{x},t)$ the density and velocity fields of the axion fluids are:

$$\rho = mB^2$$
$$\vec{v} = \frac{1}{m}\vec{\nabla}\beta \tag{16}$$

Then Eq.(14) leads to the continuity equation, and the equation of motion:

$$\partial_t v^k + v^j \partial_j v^k = -\vec{\nabla} q \tag{17}$$

with:

$$q(\vec{x},t) = -\frac{\nabla^2 \sqrt{\rho}}{2m^2 \sqrt{\rho}}.$$

Following the motion, the stress tensor is

$$T_{jk} = \rho v_j v_k + (1/4m^2)(\frac{1}{\rho}\partial_j \rho \partial_k \rho - \delta_{jk}\nabla^2 \rho). \tag{18}$$

For ordinary CDM, the last terms on the RHS of Eq.(17) and Eq.(18) are absent. In the linear regime of evolution within the horizon, neglecting second order terms, Eq.(18) becomes:

$$\delta T_{jk} = -\delta_{jk} \frac{\rho_0(t)}{4m^2} \nabla^2 \delta(\vec{x},t) \qquad (19)$$

where $\rho_0(t)$ is unperturbed density and $\delta(\vec{x},t) = \delta\rho(\vec{x},t)/\rho_0(t)$. Using conservation of energy and momentum in the background:

$$ds^2 = -[1+2\psi(\vec{x},t)]dt^2 + a(t)^2[1+2\phi(\vec{x},t)]dx^2 \qquad (20)$$

one obtains:

$$\partial_t^2 \delta + 2H\partial_t \delta - (4\pi G\rho_0 - \frac{k^4}{4m^2 a^4})\delta = 0. \qquad (21)$$

Eq.(21) implies that the axion BEC has a Jeans length:

$$k_j^{-1} = 1.02 \cdot 10^{14} \text{cm}(10^{-5}\text{eV}/m)^{1/2}[(10^{-29}\text{g}/\text{cm}^3)/\rho]^{1/4}.$$

The Jeans length is small compared to the smallest observable scales (~ 100 kpc), thus the axion BEC and CDM are indistinguishable in arena 1).

In the nonlinear regime of structure formation, the relevant equations of motion are:

$$\begin{aligned} &\partial_t \rho + \vec{\nabla} \cdot (\rho \vec{v}) = 0 \\ &\vec{\nabla} \times \vec{v} = 0 \\ &\partial_t \vec{v} + (\vec{v} \cdot \vec{\nabla})\vec{v} = -\vec{F}_g - \vec{\nabla}q \\ &q = -\nabla^2 (\psi\psi^*)^{1/2}/[2m^2(\psi\psi^*)^{1/2}] \end{aligned} \qquad (22)$$

Equation (22) is equivalent to the Schrodinger equation of particles in a Newtonian gravitational field. Axion BEC and CDM differ by the $\vec{\nabla}q$ term, which indicates a local quantum effect of axion BEC. However, as was shown by numerical simulation [8] and is expected from the WKB approximation, the differences occur only on length scales smaller than the de-Broglie wavelength.

However, we found that the gravitational interactions continues to thermalize the axion BEC after photon temperature T < 100eV$(f/10^{12}$ GeV$)^{1/2}$. The axion state thus tracks the lowest energy state. This is relevant to the angular momentum distribution of dark matter axions in galactic halos. The angular momentum distribution determines the structure and location of the caustics of galactic halos. If the initial velocity field is dominated by net overall rotation: $\vec{\nabla} \times \vec{v} \neq 0$, the inner caustic is a "tricusp ring". If the initial velocity field is irrotational: $\vec{\nabla} \times \vec{v} = 0$, the inner caustic has a "tent-like" structure [9]. Evidence has been found for tricusp rings [10], as opposed to the tent-like caustics. One can show [11] that the velocity field of non-axion cold dark matter, such as weakly interacting massive particles (WIMPs), remains irrotational, as it is the result of gravitational forces that are proportional to the gradient of the Newtonian potential. On the other hand axion BEC re-thermalizes while tidal torque is applied to it. There is a net overall rotation in this case because the lowest energy state for given total angular momentum is a state where all axions have the same angular momentum. $\vec{\nabla} \times \vec{v} \neq 0$ is accommodated through the appearance of vortices, as is observed in quantum liquids [7].

Finally, we consider the behavior of density perturbations as they enter the horizon. The axion BEC may differ from CDM because the axion BEC perturbations do not evolve linearly when they enter the horizon. The axion condensates that prevailed in neighboring horizon volumes rearrange themselves through their gravitational interactions into a new condensate for the expanded horizon volume. This produces local correlations between modes of different wave vectors. We propose this as a mechanism for the alignment of cosmic microwave background radiation anisotropy multipoles [12] through the integrated Sachs-Wolfe effect. Unlike the CDM, the ISW effect is large in axion BEC because the Newtonian potential changes entirely after entering the horizon in response to the rearrangement of the axion BEC.

We conclude that cold axions satisfy the BEC formation temperature and density conditions and thermalize by gravitational force. Although the axion BEC has no observable differences compared with CDM in the linear regime within the horizon, it provide an explanation for tricusp rings in galactic halos and may provide a mechanism for the alignment of CMBR multipoles. Although the QCD axion is best motivated, a large class of axion-like particles has the properties described here.

## ACKNOWLEDGMENTS

We thank Ozgur Erken for stimulating discussions.

## REFERENCES

1. P. Sikivie and Q. Yang Phys. Rev. Lett. 103 (2009) 111301.
2. R. D. Peccei and H. Quinn, Phys. Rev. Lett. 38 (1977) 1440 and Phys. Rev. D16 (1977) 1791; S. Weinberg, Phys. Rev. Lett. 40 (1978) 223; F. Wilczek, Phys. Rev. Lett. 40 (1978) 279.
3. J.E. Kim, Phys. Rep. 150 (1987) 1; M.S. Turner, Phys. Rep. 197 (1990) 67; G.G. Raffelt, Phys. Rep. 198 (1990) 1.
4. J. Preskill, M. Wise and F. Wilczek, Phys. Lett. B120 (1983) 127; L.Abbot and P. Sikivie, Phys. Lett. B120 (1983) 133; M. Dine and W.Fischler, Phys. Lett. B120 (1983) 137.
5. P. Sikivie, Lect. Notes Phys. 741 (2008) 19.
6. D. V. Semikoz and I.I. Tkachev, Phys. Rev. Lett. 74 (1995) 3093 and Phys. Rev. D55 (1997) 489. See also: S. Khlebnikov, Phys. Rev. A66 (2002) 063606 and references therein.
7. C.J. Pethik and H. Smith, Bose-Einstein Condensation in Dilute Gases, Cambridge University Press 2002.
8. L. M. Widrow and N. Kaiser, Ap. J. 416 (1993) L71
9. P. Sikivie, Phys. Rev. D60 (1999) 063501.
10. L.D. Duffy and P. Sikivie, Phys. Rev. D78 (2008) 063508), and references therein.
11. A. Natarajan and P. Sikivie, Phys. Rev. D73 (2006) 023510.
12. M. Tegmark, A. de Oliveira-Costa and A. Hamilton, Phys. Rev. D68 (2003) 123523; A.de Oliveira-Costa, M. Tegmark, M. Zaldarriaga and A. Hamilton, Phys. Rev. D69 (2004) 063516; C.J. Copi, D. Huterer, D.J. Schwarz and G.D. Starkman, MNRAS 367 (2006) 79 and references therein.